\documentclass{llncs}
\usepackage[ruled,vlined]{algorithm2e}
\usepackage[utf8]{inputenc}
\usepackage{glossaries}
\usepackage{parskip}
\usepackage{hyperref}
\usepackage{graphicx}
\usepackage{listings}
\usepackage{appendix}
\usepackage[table,xcdraw]{xcolor}
\usepackage{hhline}
\usepackage{float}
\usepackage[T1]{fontenc}

\usepackage{color}

\definecolor{pblue}{rgb}{0.13,0.13,1}
\definecolor{pgreen}{rgb}{0,0.5,0}
\definecolor{pred}{rgb}{0.9,0,0}
\definecolor{pgrey}{rgb}{0.46,0.45,0.48}

\usepackage{listings}
\title{The Requirement Gatherers' Approach to the 2019 Multi-Agent Programming Contest Scenario}
 \author{Michael Vezina\inst{1}\orcidID{0000-0001-8109-6486} \\ Babak Esfandiari\inst{1}}

\institute{Carleton University, Ottawa ON, Canada}

\date{April 2020}

\begin{document}

\maketitle

\begin{abstract}
    The 2019 Multi-Agent Programming Contest (MAPC) scenario poses many challenges for agents participating in the contest. We discuss The Requirement Gatherers' (TRG) approach to handling the various challenges we faced --- including how we designed our system, how we went about debugging our agents, and the strategy we employed to each of our agents. We conclude the paper with remarks about the performance of our agents, and what we should have done differently.
\end{abstract}

% Introduction
\section{Introduction}
Each year, the Multi-Agent Programming Contest (MAPC)\footnote{\url{https://multiagentcontest.org/}} releases a scenario that discusses the specifics of the contest. The scenario goes into depth about the simulation environment, the perceptions and actions available to each agent, and the requirements that the agents must satisfy in order to perform well in the contest.

We introduce the paper with a brief description of the 2019 MAPC scenario. We then go on to discuss the specifics of the The Requirement Gatherers' (TRG) team, starting with the design challenges and motivations that drove the main system design decisions. The design decisions that were made are then discussed in detail, including the purpose of each of the individual system components, the interactions between them, and the purpose they serve in the system. The various setbacks encountered with debugging our multi-agent system are discussed, along with the various approaches used. We introduce a visualization tool that helped significantly with debugging our agents when used in tandem with the standard agent debugging tools. 

The team strategy section will look at the main behaviour and strategy of the various agent roles, and what part each of them play in the overall simulation. We then conclude the paper with some remarks about TRG's performance in the competition, using the gathered agent metrics to help speculate why the team performed the way that it did.  

% Background: Understanding the Scenario
\section{The 2019 Scenario: Agents Assemble}
The MAPC scenario varies from year to year, this year (2019) the scenario is named ``Agents Assemble''. Two teams are placed on a map represented by a rectangular grid and are required to complete tasks by their respective deadlines to compete for the most points. Each agent is provided with a list of tasks to be completed, and can collaborate with their team to determine which task to complete and how a given task can be completed. 

The scenario places a lot of limitations on the agents, including a (severely) limited perception range, lack of agent identification perceptions, events that destroy and regenerate areas of the map, among other additional limitations. We will discuss all of the limitations and challenges faced, and our solutions to them, in detail throughout this paper. 

% System Design
\section{System Design: Challenges and Motivations}
Jason\footnote{\url{http://jason.sourceforge.net/}} \cite{JasonBook} is an AgentSpeak(L) \cite{AgentSpeak} interpreter built on Java that was made specifically for the development of multi-agent systems. Jason provides various agent-oriented features on top of its ability to interpret AgentSpeak code. Jason does all of the heavy-lifting with regards to how the agents reason, while also allowing the agent developer to customize any aspect of the architecture. On top of this, Jason also provides the ability to specify an agent environment, customize various selection functions, and allows for the calling and exchanging of information to (and from) Java code through the usage of internal action functions. 

The general system design approach for this team is to use AgentSpeak code for the implementation of agent behaviour and strategy, using Jason to provide its powerful interpreting, reasoning, and customization abilities. The AgentSpeak code will utilize Jason's internal actions to defer to Java code when it is necessary or more convenient to do so. This section discusses some of the design concerns that were encountered during the process of agent development.

\begin{figure}[h]
\centering
\includegraphics[scale=0.55]{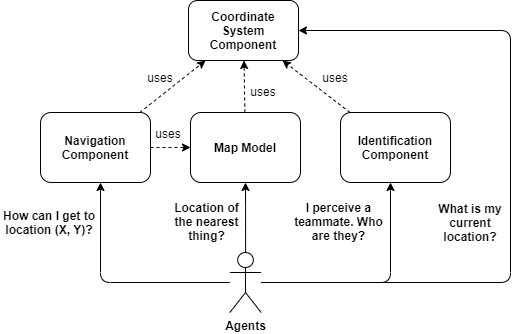}
\caption{A high-level diagram showing some of the main system components. The dependencies of each component are shown with dashed lines, and the questions that each agent may ask the component are in bold. The Agents actor is meant to demonstrate how the behaviour of the agent (implemented in AgentSpeak) interacts with these main components to form the basis of its strategy.}
\label{fig:component-diagram}
\end{figure}

We will briefly describe the approach that our team took to break down the different challenges faced throughout the course of the competition. Figure \ref{fig:component-diagram} shows a high-level diagram detailing the main system components and how the behaviour of our agent interacts with these components. The usage of these high-level components by the agent with respect to its specific goals and plans will be discussed in detail in the team strategy section.

Our team of agents utilizes a coordinate and navigation system with a data structure --- known as the map model --- that stores the map knowledge of each agent as they perceive the map.  These components are the main building blocks that allow the agents to navigate and reason about their map knowledge. The agents also utilize the coordinate system as a basis for unique agent identification. As each agent moves around the map and perceives other team agents, they will attempt to determine the identity of one another. Identification allows the identified agents to collaborate and share information with each other. The updating of the map model, and the agent identification process, occurs asynchronously from any other tasks being carried out by the agent. 

In the team strategy section, we demonstrate how the agents can then use the high-level system design components to fulfill their immediate goals and ultimately complete tasks in the simulation. Agents who work on the tasks are called \textit{builders}. The builders are assigned tasks in sub-teams. One builder in the sub-team is assigned the status of the \textit{master builder}, and acts as a centralized point to coordinate, connect, and submit the requirements. The other builders (the \textit{slave builders}) are responsible for obtaining their respective blocks from dispensers, and then delivering and connecting their block to the master builder. Once all task requirements have been connected, the master builder can submit the task on a goal space.

Although the navigation, coordinate, map model, and identification components do not constitute the complete system, we demonstrate the purpose of these main components and how the agents interact with them in Figure \ref{fig:component-diagram}. This is a high-level diagram showing the main questions that each component can answer.

Some of the design challenges that we faced stemmed from certain limitations in Jason and multi-agent system development in general; these are discussed in the general software engineering design concerns. The other concerns we faced were due to intentional limitations put in place by the MAPC scenario; these will be discussed separately from the general software engineering design concerns. Each section will discuss its relevant design concerns and motivations, and will provide a list of design requirements in order to address the concern at hand.

\subsection{General Software Engineering Concerns}
The general software engineering concerns include some of the limitations that were encountered with agent development; these concerns and motivations are independent of the contest scenario limitations. The concerns discussed in this section include the following:
\begin{enumerate}
    \item Requiring access to readily-available agent state information from outside the agent code
    \item Ensuring a proper separation of concerns between the agent code and any operations delegated to outside of the agent code
    \item Ensuring proper synchronization between the agent reasoning cycle and the simulation cycles
\end{enumerate}

\subsubsection{Agent State Information Access}
Components such as the map model and navigation system constantly need access to agent-specific state information, specifically the agent's map perceptions, outside of the AgentSpeak code. These components are accessed by the agents through internal actions. One method of providing agent-specific state information to the components is to retrieve the necessary information from the provided agent's transition system. The transition system is part of Jason's architecture and provides access to various agent information, including the current state of its perceptions and beliefs.

Unfortunately, the transition system does not provide a hassle-free way to quickly access state information. More often than not, iteration and type-casting are required before you are able to make any use of the information provided by the transition system. This is demonstrated in Listing \ref{lst:transition-system}, where we show how the transition system can be used to obtain and process information about all block perceptions. 

The additional overhead associated with using the transition system is necessary in order to perform any processing on the current agent state perceptions. Every time an internal action is called, the code shown in Listing \ref{lst:transition-system} is executed.  Internal actions can be invoked many times during the deliberation stage; requiring iteration and type-casting every time they get called. This can have a significant impact on agent performance, reliability, and modifiability. We would like to introduce the following design requirements to address this concern:

\begin{enumerate}
    \item A container for the current state of each agent that exposes an interface allowing for the querying of agent state information. The amount of type-casting required in order to process the information provided by this container should be minimized.
    
    \item This container should process any new perceptions only once per simulation step.
    
    \item A way to access the above-specified containers. This should be accessible by both the internal action classes and the environment class.
    
    \item In order to further improve agent performance, the parsing and processing of percepts done by the container should be done on its own thread to minimize the impact on the agent threads.
\end{enumerate}

\begin{lstlisting}[language=Java, caption={Accessing agent state information from the transition system within an internal action. Most of this is boilerplate code.}, label={lst:transition-system}]

Literal desiredPercept = ASSyntax.parseLiteral("thing(X, Y, block, Details)");

// Use pattern matching to find the desired perception
ts.getAg().getBB().getCandidateBeliefs(desiredPercept, un)
    per -> {
        // Return if not a percept (no percept source)
        if(!perLiteral.hasSource( ASSyntax.createAtom("percept")))
            return;
            
        // Only way to retrieve percept info.
        // Smelly: type-casting and term indices.
        NumberTerm xTerm = (NumberTerm) per.getTerm(0);
        NumberTerm yTerm = (NumberTerm) per.getTerm(1);
        Term typeTerm = per.getTerm(2);
        Term detailsTerm = per.getTerm(3);
        try {
            // Resolve and cast the coordinates
            int x = (int) xTerm.solve();
            int y = (int) yTerm.solve();
            
            // Obtain block details
            String thingType = typeTerm.toString();
            String details = detailsTerm.toString();
        
            // Process the percept information
            processBlock(x, y, details);
        } catch (NoValueException noValEx) 
        { ... } // NumberTerm.solve() failed 
    });
\end{lstlisting}

\subsubsection{Separation of Concerns}
The ability to call custom internal action functions within the AgentSpeak code in Jason allows the agents to utilize Java to implement some unit of functionality for the agent. The internal actions have access to all agent information, allowing the internal actions to make any necessary calculations or decisions based on the current agent state.  However, it can be easy to rely on the internal actions to implement agent behaviour or logic that should typically be done through AgentSpeak. In this case, you may be losing out on the following potential benefits of Jason.

Jason provides various mechanisms for agents in a dynamic environment. The ability to specify contingency plans in Jason allow the agents to manage, or even rectify, failures or unexpected states caused by any agent, or the environment. As part of each reasoning cycle, Jason will ensure that the agent is performing its most desirable intentions --- taking the current agent beliefs and environment perceptions into account. This means that the agents may drop its current intentions at any point in time, in order to pursue a goal or desire of higher importance. On top of this, Jason is inherently event-based, which allows the agent to manage events and perceptions as they are received from the environment.

The moment that strategy is implemented using internal actions rather than being expressed through Jason plans, the agents lose out on Jason's innate ability to re-evaluate the agent's goals and handle other current events occurring within the dynamic environment. Internal actions are treated as atomic operations with respect to the reasoning cycle. They have their place, which is to process information and provide a result, provide the agent with information not available through percepts (for example, using the map model to obtain data from outside the agent's perception range), or to provide a way of interfacing with higher-level components (such as utilizing the navigation system's path-finding algorithm).

By maintaining a proper separation of concerns, we can ensure that the reasoning mechanisms provided by Jason are being used to their full potential, while also making appropriate use of internal actions when necessary. Our agents attempt to use their immediate perceptions as much as possible, only deferring to internal actions when it is more appropriate or convenient. 

This can be seen in the case of agent navigation. Agents will use the navigation system to find something on the map. However, since the map model used by the navigation system may be out of date, the path returned by the navigation system may not be accurate. The agent still utilizes the path, but uses its immediate perceptions to confirm whether or not the returned path is blocked. During navigation, if the immediate perceptions contradict or block the path returned by the navigation system, the agent will request a new path.

\subsubsection{Reasoning Cycle}
For the competition, it was desirable for the agents to perceive, think, and act asynchronously from the server. This means that an agent is allowed to perceive and deliberate multiple times per simulation step, until it decides on an action to perform. Once an action is performed and sent to the server, the agent waits until the beginning of the next simulation step before continuing onto its next reasoning cycle. Having the agents perceive and deliberate asynchronously from the simulation allows the agents to process multiple aspects of their current environment.

By contrast, if the reasoning cycle was synchronized with the simulation step this would introduce contention between the various events that need to be processed by the agent; only one event can be processed per simulation step. This means that the agent must choose between handling incoming messages from other agents, handling belief base events (such as the addition and removal of beliefs), and executing its current intentions. All of these events must be processed within the span of a simulation cycle, therefore driving the requirement for having the agents reason asynchronously from the simulation.

\subsection{MAPC Scenario-Specific Design Concerns}
This section discusses some of the design concerns introduced by the 2019 ``Agents Assemble" scenario. These concerns stem from the limited perception range and map information provided to agents, randomized map destruction and regeneration (clear events), not being able to uniquely identify teammates through perceptions, and not being able to rely on the provided perceptions to determine the current agent's attachments. 

Each of the following sections will discuss the challenges imposed by the scenario, and the design requirements that need to be met in order to bridge some of the gaps and provide a system that allows agents to reason and act as reliably as possible in their environment.

\subsubsection{Navigation Challenges}
The simulation provides the agents with no information about their current location, no information about the map boundaries, and, on top of this, the simulation also restricts the perception range of the agent's map surroundings. The limited perceptions that are given to the agent are provided relative to the agent's location. 

\begin{figure}[h]
\centering
\includegraphics[scale=0.55]{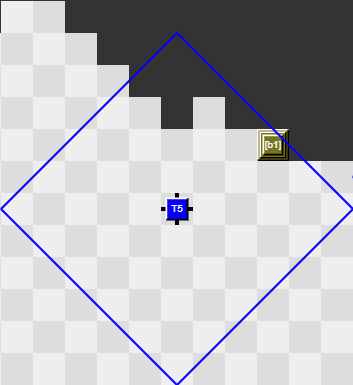}
\caption{The limited perception range of each agent. The agent only receives perceptions from within the blue bounding box. This image was captured from the simulation monitor.}
\label{fig:perception-range}
\end{figure}

This makes navigation not only difficult, but also unreliable, due to each agent having a restricted view of the map and the fact that the map is extremely dynamic in nature. The agents use a path-finding algorithm in order to provide best-effort navigation. The agents must have some contingencies in place, just in case the dynamic environment causes the path-finding algorithm to provide a non-navigable path. Figure \ref{fig:perception-range} shows the limited perception range of each agent, limited to within the blue bounding box.

In order for the agents to navigate as reliably as possible, the following requirements must be met:
\begin{enumerate}
    \item Path-finding can only provide a path on the current map knowledge. We must maximize the agent's amount of up-to-date map knowledge to provide the best possible navigation path.
    
    \item Path-finding typically requires the map be represented as a graph, with nodes and edges representing the possible paths the agent can take. This means that map knowledge must be processed and provided as a graph, and that the interaction between an agent and the map elements must be properly modelled through the edges between navigable nodes.
    
    \item Map knowledge is constantly updated, and so adding new map knowledge to the graph should be as quick as possible. The graph should be initialized at the beginning of the simulation, and updated every new simulation step.
    
    \item Since the path-finding algorithm will have to rely on outdated information outside of the agent's perception range, agents must be able to handle cases where the path given by the path-finding algorithm is blocked or inaccurate. The agent must be able to detect when the provided path is no longer valid and must be able to handle this.
\end{enumerate}

\subsubsection{Coordination Challenges}
Agent coordination, which includes information sharing and task collaboration, is also made complicated by the simulation. On top of not being able to access any form of absolute positioning, agent teammates are also not uniquely identified by perceptions. To demonstrate, if the teammates agent T1 and agent T2 are within each other's range of perception, neither agent will receive any information regarding the identity of the other agent, except for the fact that they are teammates. This means that agent T1 does not know if the agent it is perceiving is agent T2, or if it is a different teammate (maybe it is agent T3). This becomes an issue specifically when agents are required to collaborate on tasks in the simulation.

In order to connect blocks together to satisfy task requirements, agents must use the \textit{connect} action. The connect action allows two agents to connect two blocks together. When using the connect action, each agent must specify the name of the other agent it wishes to connect blocks with. Due to this, the agents must uniquely identify each other in order to perform any level of collaboration. 

Although not a necessity, it would be in our best interest if agents could utilize identification to enable information sharing. Sharing information between agents would ensure that agents could collaborate more consistently and effectively, and would also require less time exploring the map since they could combine their map knowledge. In order to communicate knowledge between two agents, they must be able to uniquely identify each other while also providing a basis for communicating relative information (specifically, the map knowledge for each agent). In order to achieve agent coordination, the following requirements must be met:
\begin{enumerate}
    \item There must be a mechanism to uniquely identify teammate agents
    \item There must be a way to share information (including location information) between agents without the need for absolute coordinates from the simulation
\end{enumerate}

\subsubsection{Attachment Monitoring}
The attachment perceptions provided by the simulation are generic and unreliable for identifying which agent the thing (such as a block or another agent) is attached to. For the sake of simplicity, the agents will only care about attached blocks (and will not pay any attention to any agents attached to each other). 

The attached percept only provides information about the relative X and Y coordinate of the attached block and will not identify the agent that is attached to the block --- even if the block is attached to the current agent. This means that agents must put further measures in place in order to keep track of the blocks they have attached to themselves.

The agents must be able to reliably keep track of their current attachments, as it allows them to determine whether or not they have the required blocks to complete tasks. Through attachment monitoring, the agents can detect if they lose an attachment and handle this scenario as appropriately and quickly as possible. Attachment monitoring also helps to accurately determine the movement and rotation directions that are blocked by attached blocks. This is used in navigation to determine which movement and rotation directions are unblocked. This ultimately helps reduce the amount of steps an agent may waste trying to move or rotate in a direction that is blocked by an attachment.

\subsubsection{Requirement Planning}
The requirements for each task in the simulation must be connected in a sequential order, otherwise it becomes impossible for agents to connect some of the requirements. The simulation does not provide the task requirements in any specific order, so we require an algorithm that plans out the task requirements, allowing the agents to connect the blocks in the required order.

\begin{figure}[h!]
\centering
\includegraphics[scale=0.5]{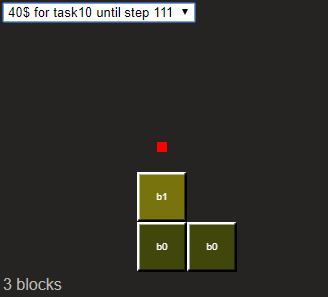}
\caption{An example of a task generated by the simulation. The red square represents the location of the agent, with respect to the attachments. This image was collected from the simulation monitor.}
\label{fig:l-shaped-task}
\end{figure}

Figure \ref{fig:l-shaped-task} shows an L-shaped task. In the case of the figure, the requirement planning algorithm would have to return the sequence as follows: requirement b1 at (0, 1), requirement b0 at (0, 2), and then requirement b0 at (1, 2). Looking at this figure, it is easy to see that it would be impossible to connect the b0 block at (1, 2) before the b0 block at (0, 2).

As we discuss the overall system design and how it supports our team strategy in the upcoming section, we utilize these design concerns and motivations, and the introduced design requirements, to justify the main system design decisions that were made.
\section{System Design}
The design of the system aims to address both the software engineering and MAPC scenario-specific design concerns mentioned in the previous section. We will introduce the various components of the system, how they interact with each other, and how they solve the various design concerns. The team strategy and agent behaviour implemented in AgentSpeak will be discussed in the team strategy section. This section mainly discusses the high-level system components that provide a basis for the agent behaviour to build off of. The components discussed in this section are built in Java, and are accessed by the agents through the usage of internal actions.

\subsection{General System Design}
\begin{figure}[h!]
\centering
\includegraphics[width=0.85\textwidth]{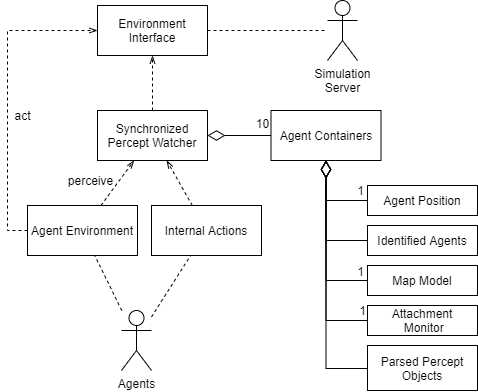}
\caption{The structure of each of the high-level components and how the agent interacts with them. The strategy of the agents, which is discussed in detail in the team strategy section, is represented by the Agents actor.}
\label{fig:component-structure}
\end{figure}

This section discusses the main components and their interactions of the system. These components interact with one another in order to address the general software engineering concerns as a whole. The components in this section are the synchronized percept watcher, the agent containers, and the parsed percept objects. Figure \ref{fig:component-structure} shows the general structure of each of these components, and Figure \ref{fig:sys-flowchart} shows the interaction between these system components. These system components process and parse the raw percepts provided by the simulation into usable and readily-accessible information through the usage of various objects. As a whole, these components work together to make information access straight-forward, while also doing it in a way that minimizes the performance impact on the agent threads.

\subsubsection{Synchronized Percept Watcher (SPW)}
The synchronized percept watcher (SPW) is one component responsible for addressing some of the design concerns stated in the previous section. The SPW is a layer implemented in Java and sits between the competition-provided environment interface (EI) object (which communicates with the server directly), and the agent threads. It runs on its own thread, and polls the EI object for new agent percepts. Upon receiving new percepts, the SPW will parse and process them into containers for each agent, appropriately named \textit{agent containers}. 

The SPW was designed to be thread-safe; any attempts to access an agent container before new percepts have been completely processed will be blocked until the container is ready. In most cases, the SPW thread can parse and process the perceptions before the agent threads attempt to use the containers. The SPW parses and processes the new percepts only once per simulation step, caching the resulting agent containers. The agent threads can access and utilize these containers as much as they need, without requiring the containers to re-process the information provided by the server. The SPW provides global access to the agent containers through its singleton instance. This means that both the agent environment class and the agent internal action classes have access to the current agent state through the containers.

\subsubsection{Agent Containers}
The agent containers were developed to provide a single object that contains all information about an agent. Each agent container contains the agent name, the current step percepts, the current location of the agent, the complete map model object, the collection of currently identified teammates, and various other objects. The agent containers are created and updated by the SPW. Upon every new set of perceptions, the SPW will call each agent container, passing in the new set of raw perceptions from the server. The container will parse the percepts and update all relevant information about the agent (such as the current virtual location of the agent on the map). Each agent container also maintains the raw perceptions received by the server so that they may be directly provided by the environment to the agent during its perceive cycle. The agent containers also provide access to higher-level information and processing on the percepts. This information and processing is only available to the agents through internal actions. 

This can be seen with the map model contained within the agent container. The map model contains a representation of all perceived map information over the course of the simulation --- this includes perceptions that are currently outside our vision range. The map model also allows us to perform complex operations on the data, such as path-finding and map synchronization with identified teammates. Any information stored in the agent container, such as the map model object, needs to be requested and transferred through internal actions.

\subsubsection{Parsed Percept Objects}
The agent container also contains a collection of various percept classes. For every percept an agent may receive, there is a corresponding Java object. Each raw percept received by the SPW is parsed into their relevant Java objects by the agent container. The percept classes were developed with the purpose of containing the parsed information from each percept, while also providing appropriately-named get methods to retrieve the information. For example, it is possible to obtain perceived terrain information through the agent container and the parsed percept objects without needing to find and parse the perception from a generic Literal object (as is required when accessing perception information through the transition system).

\begin{figure}[h!]
\centering
\includegraphics[width=0.7\textwidth]{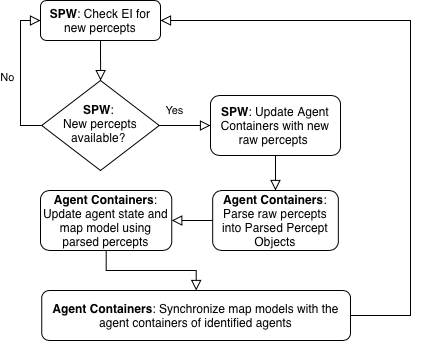}
\caption{A flowchart showing the general flow of information parsing and processing in the system, including how the agent containers are updated by the SPW.}
\label{fig:sys-flowchart}
\end{figure}

Figure \ref{fig:sys-flowchart} shows how all of the system components interact with each other in order to update the agent state contained within the agent containers. The agent containers can be accessed from anywhere within the internal action or environment classes through the SPW singleton instance.

\subsection{MAPC Scenario-Specific System Design}
This section will discuss the design approach to dealing with all of the navigation requirements. This includes how the agents manage the limited perception range, incorporating a positioning system that can be used and understood by the agents, how map information is gathered and processed so that it may be usable by the navigation system, and how agents can use map information to intelligently explore the map and utilize the path-finding algorithm.

\subsubsection{Virtual Absolute Positioning System}
To help the agents make sense of their surroundings, we introduced a virtual positioning system. Each agent starts off with an initial position of (0, 0), regardless of their actual location in the simulation. Successful move actions are used to update the agent's current location. This system provides an absolute location based off of the reference point of (0, 0). Providing the agents with this positioning system allows them to keep track of map elements that are no longer part of their immediate perception range. The positioning system also helps form the basis of path-finding, since they can express destinations using virtual positions. Each location in the positioning system corresponds to one cell location on the simulation map. 

The current virtual position of each agent is stored and updated by their respective agent containers. In order to provide the position information to the agents, the environment class will obtain the position from the container during the agent's perceive stage. Once obtained, the environment will provide the virtual position as a perception to the agent. The virtual position perception is one of the few pieces of information that is actually provided to the agent. This is because agents must be able to reason about their location, including determining whether or not path-finding is required.

\subsubsection{The Map Model}
Now that the agents can utilize the virtual positioning system in order to remember their surroundings as they navigate the map, we must create a structure that allows the agent to store and query the perceived map data. This map data must also be accessible in the form of a graph in order to be used by the A* algorithm. This is where the map model comes in. Each agent has their own map model, as it builds off of the virtual positioning system. The map model contains the graph object used for path-finding, and a list of current map perceptions.

After the agent containers parse the current perceptions, the container packages them up into corresponding \textit{MapPercept} objects. Since it is possible to have multiple percepts corresponding to the same cell location on the map, the map model utilizes the MapPercept object to contain all percepts about a given cell. The map model creates one MapPercept object for each cell location in the perception range of the agent, and then places the perception information into the appropriate MapPercept object. Each MapPercept object contains the virtual position, the perceived terrain information, and a list of perceived things for its corresponding cell. The MapPercept object provides methods that determine how the current agent interacts with the terrain and things on that cell. These methods are utilized by the path-finding graph to model the possible agent movement with edges.

The map model contains all of the map data ever perceived by the agent. Since the positioning system is relative to each agent, it would not be possible for team agents to utilize the knowledge in each other's map model. This process of map model synchronization is possible once agents have uniquely identified each other, and can provide a way to translate knowledge between the two agent's coordinate systems.

\subsubsection{Agent Identification}
As a crucial component of team coordination, agents are required to identify one another. One of the many restrictions of the simulation is that agents can not perceive the specific identity of the agent. For example, assume we are agent T1 and we perceive two teammates: agent T2 and agent T3. The perceptions provided to agent T1 specify the relative location and the team that agent T2 and T3 are on, but do not specify which perception corresponds to agent T2 and which corresponds to agent T3. This, in tandem with not knowing your absolute position in the simulation, makes it difficult for two agents to communicate directly and ultimately poses a major challenge for precise and reliable team coordination.

The process of agent identification starts off with each agent having no knowledge of other teammates. Upon perception of a team member, they will attempt to identify one another through a simple process that utilizes their relative coordinates. Upon perception of a teammate, both agents will broadcast the fact that they perceive a teammate to the operator agent.

It is now the job of the operator to determine whether or not the identity of a teammate perception is trivial. Agents perceive each other in pairs, if agent T1 perceives agent T2, then agent T2 will always perceive agent T1. The operator can use the perceived relative locations to deduce the identity of each agent in the pair, based off of the agents that broadcasted a message to the operator. If more than one pair of agents perceive each other at the same relative location, the operator is not able to deduce the identity of any of the agents in the pairs. If this is the case, the operator abandons the identification process, and the agents continue as if nothing happened. This process repeats every time agents perceive each other, so if identification fails, it is likely that the agents will eventually be able to identify one another in the near future. Once agents identify each other, they will be identified for the remainder of the simulation.

Upon successful identification between two agents, say agent T1 and agent T2, each agent calculates the \textit{translation coordinates}. The translation coordinates correspond to the (X, Y) value that represent the difference between agent T1 and agent T2's virtual location of (0, 0). The translation coordinates provide a way for identified agents to communicate virtual positions to one another, including meeting points (for task building and submission) and for sharing map information with each other, allowing them to synchronize their maps and contribute to a common map model. 

\subsection{Monitoring Agent Attachments}
Since each agent must keep track of their own attachments, and are not able to rely on attach percepts, we must add a mechanism that keeps track of attached blocks in an intelligent way. This mechanism should update its model any time the agent attaches, detaches, or connects a block, which should be a simple task. However, it is also possible for blocks to be destroyed by clear events and task submissions.

To take care of this, we can create an internal model for each agent that contains a collection of all of the blocks attached to the agent. One thing to note, is that we make the assumption that we will never attach to another teammate. Although this is possible in the simulation, the agents proactively ensure that this never happens. Additionally, the agents ensure that they only attach and connect one block at a time. This makes the attachment model logic simple, yet accurate.

Any time the agent attaches, connects, or detaches a block, the agent updates the model through an internal action. To account for clear events, and task submissions, we allow the attachment model object to refresh itself at the start of every step. The model can utilize the absence of an attach percept to determine if we have lost an attachment. If the model thinks it has an attachment, but the simulation does not provide an attach percept, the model can safely remove the attachment. This handles the impact of an unexpected clear events and task submissions, and allows the model to maintain an accurate representation of what blocks are attached.

\subsection{Task Requirement Planning}
Each task is defined with the list of requirements necessary for successful task submission. Each requirement defines the block type and the required attached location of the block, relative to the submitting agent. In order to ensure that all requirements can be connected to the submitting agent, the requirements must be planned out. 

\subsubsection{Understanding Task and Requirement Generation}
Each requirement is composed of an (X, Y) coordinate, and a block type. In order to ensure that agents can connect the generated task requirements, the simulation server generates the requirements in a specific sequence. We can utilize the same rules that the server uses to generate the requirements in order to plan out the proper sequence of requirement connections. The coordinate of the first requirement will always be (0, 1), this is directly south of the attaching agent. Every requirement after that can only be south, east, or west of the previous requirement location. The simulation will never generate a requirement north of its predecessor. 

\begin{figure}[h!]
\centering
\includegraphics[width=\textwidth]{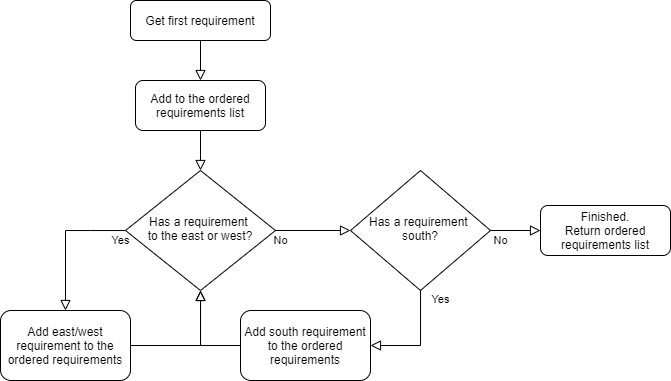}
\caption{A flowchart demonstrating how the ordered sequence of requirements is built from an unordered list.}
\label{fig:req-flowchart}
\end{figure}

We can exploit these rules in order to come up with the requirement sequence. We search through the list of unordered requirements looking for the requirement with a location of (0, 1). Once found, we look for a requirement to either the east (1, 1) or west (-1, 1) side, and if one does not exist, we look south (0, 2) for a requirement. We then repeat this for every requirement until we exhaust the list of requirements. The end result is an ordered list of requirements that the builder agents can use to build the tasks in the correct order. Figure \ref{fig:req-flowchart} demonstrates this procedure through the usage of a flowchart.

% Debugging the System
\section{Debugging}
Most of the development time was spent on debugging unwanted agent behaviour. In some cases, the debugging tools that were used added more resistance to development and debugging efforts. One tool specifically --- the agent mind inspector, which is a debugging tool provided by Jason --- was very helpful during the initial stages of development, where there was only one agent in operation and the single agent had very simple behaviour. As soon as additional agents were introduced and the behaviour became more complex, debugging the agents using the mind inspector started to become a time-consuming and troublesome task. We needed to find a better solution to debugging the agents. This section will discuss the debugging techniques that were used during the course of agent development. This includes debugging efforts that involve the Jason mind inspector, the debugging support provided by the IntelliJ IDE\footnote{\url{https://www.jetbrains.com/idea/}}, a custom-built agent state visualization tool, and some combination of all of the tools.

\subsection{Jason's Mind Inspector}
The Jason mind inspector provides the ability to debug and step through the behaviour of the agent. Unfortunately, debugging the behaviour of multiple agents using the mind inspector became a troublesome task as behaviour increased in complexity. When debugging using the Jason mind inspector, you are able to pause and step through the execution of one or all of the agent reasoning cycles. Since the simulation executes separately from the agent threads, the simulation will continue its execution even though the agent threads are paused for debugging. This causes synchronization issues between the agents and the simulation, which in turn introduces other issues, making it difficult to trace down the root cause of the original bug. This is not an issue specific to Jason, or even multi-agent systems, but rather an inherent issue in debugging multi-threaded applications.

In an attempt to combat the synchronization issue, the mind inspector's ``history" option was used. This option allows the inspector to keep a full history of all agent reasoning cycles. Using this option allows the developer to keep the agent threads running, while still providing a way for the developer to examine the recorded reasoning cycles and behaviour. Unfortunately, keeping a full history of every piece of information for each reasoning cycle quickly grew out-of-hand, causing the debugger to be sluggish and completely unusable. To make matters worse, the debugger would also slow down all of the agent threads, making the inspector's history option a dead-end. Breakpoints in Jason allow the debugger to run until a breakpoint (specified in the AgentSpeak code) is hit. Unless the ``history" option is enabled, it is extremely difficult to determine the context of the current reasoning cycle, since you do not have a trace of how the agent got to the current state.

Since the debugging tools provided by Jason were designed for the specific purpose of debugging agent behaviour, they were limited to only being able to see the information available in the AgentSpeak code. If any bugs occurred with any of the Java components (such as the agent's container or the map model), the mind inspector would not be able to provide any insight. This is where IntelliJ's debugging support comes in.

\subsection{IntelliJ IDEA IDE Debugging Support}
IntelliJ, which is a Java IDE, provides some excellent tools and support for debugging Java applications. We can utilize the debugging tools provided by IntelliJ to help us debug our system. Debugging issues with any of the Java components was straight-forward, all that was required was setting a breakpoint and stepping-through the operation of the component. This method of debugging is still prone to the multi-threaded debugging issues, although this issue can be alleviated through the usage of conditional breakpoints. This allows us to debug the code when necessary (and only under certain conditions), and so getting out-of-sync with the simulation only occurs when the breakpoint gets triggered. At this point, if the agents get out-of-sync with the simulation, it typically doesn't matter since we have the relevant information required to examine the issue from the agent containers.

Although this method of debugging was less troublesome than the mind inspector, there was still a large amount of information that had to be processed in order to diagnose any issues. For example, there were initial issues with the virtual positioning system, but they did not express themselves until the agents had to perform path-finding on the map model. Because of all of the interactions between the components, this issue was very difficult to diagnose, even with IntelliJ's debugging support. On top of this, IntelliJ does not provide any support for debugging the agent's reasoning cycle.

\subsection{Agent Visualization Tool}
In order to help reduce the amount of effort that went into debugging, a custom visualization tool was created. This tool provides a visual representation of each agent's current state, providing full flexibility in the information that the tool can display. This tool provided a way to monitor the state of the agent. When any abnormalities were detected in either the behaviour of the agent or the information displayed by the tool, it was easy to pin-point the component that was causing the issue.

For example, the previously-mentioned issue relating to the virtual positioning system was quickly detected with the visualization tool. By having the visualizer show the agent's knowledge of the map as well as the current location of the agent, it was obvious that the virtual position was not updating in the correct circumstances --- in this case, it was not updating correctly whenever the agent attempted to move into an obstacle. Without the visualization tool, a lot of breakpoints and stepping through execution would have been required to determine the root cause. The visualization tool shows everything about the agent: map information, the list of identified agents, the current navigation path of the agent, the current attached blocks, and more. 

\begin{figure}[h!]
\centering
\includegraphics[width=\textwidth]{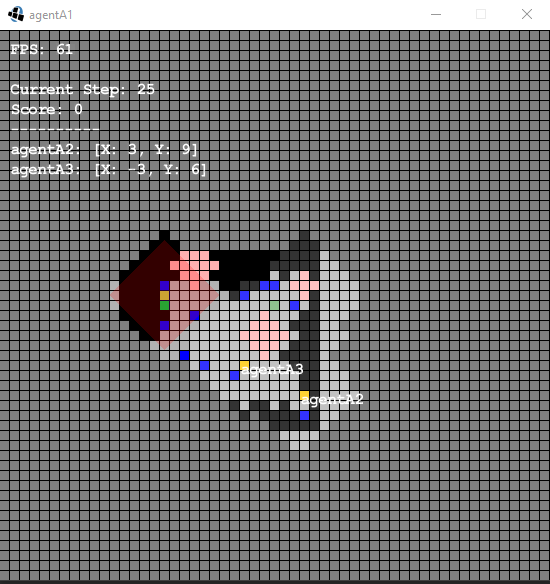}

\caption{A demonstration of the information that can be displayed through the visualization tool.}
\label{fig:visualization-tool}
\end{figure}

Figure \ref{fig:visualization-tool} shows the information conveyed by the visualization tool. The figure shows the map knowledge as a collection of square cells. The different cell colours represent the various things or terrain. To name a few examples, the yellow cells are team agents, black cells are obstacles, pink cells are goal spaces, and the blue cells are block dispensers. The dark grey cells are the locations that the agent has not explored yet. The figure also shows the current simulation step information at the top left corner, along with a list of the team agents that have been identified by the current agent (and their current location on the map). Information can be added to and removed from this tool, in order to support the requirements of the developer. For example, it is possible to add in information about the agent's current goals and intentions. 

\subsection{The Combined Approach}
In the end, a combination of all three debugging approaches were used. The visualization tool was used to monitor both the Java and AgentSpeak components. Any time unwanted behaviour occurred, the visualization tool would provide enough information to create conditional breakpoints in IntelliJ, reducing the amount of guessing work and time required to pin-point the root of the bug. The visualizer was also used for debugging the behaviour implemented in AgentSpeak. Any time any strategic or behavioural issues were visually detected through the tool, we were able to invoke the Jason mind inspector at that moment in time in order to examine how the agent is reasoning. This tool ultimately reduced the amount of guess work involved in debugging the agents, and improved the debugging experience as a whole.

% Team Strategy, Discussion of Agent Deliberation
\section{The Team Strategy}
The team is composed of an operator and two main roles: the attackers, and the builders. The operator is a single agent that acts as a centralized point for assigning task requirements and processing agent identification notifications from other agents. The operator does not connect or perform in the simulation, but does communicate with the other agents. Out of the agents that do perform in the simulation, each agent is assigned either a role of attacker (also known as an offender), or builder. During the competition, agents T1 to T5 were builders, while agents T6 to T10 were assigned the attacker role. The attackers are agents that explore the map with the sole purpose of interfering with the other team. The builders are the agents who collaborate with one another in an attempt to build and submit tasks.

\subsection{The Operator}
The operator is an agent who does not participate in the simulation. Instead, the operator agent acts as a centralized point for processing team information, such as processing agent identification messages, monitoring currently assigned tasks, and assigning new tasks to \textit{free agents}. The operator's role in agent identification has been previously discussed in the system design section, it simply just acts as a centralized point for organizing identification broadcasts into pairs based off of the relative location of perceptions. The operator then defers to an internal action to carry out further identification logic. This section focuses on the operator's role in selecting and assigning tasks to sub-teams of builder agents. 

\begin{figure}[h!]
\centering
\includegraphics[width=0.8\textwidth]{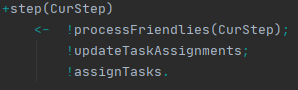}
\caption{The operator's plan for handling new simulation step events.}
\label{fig:operator-plan}
\end{figure}

Figure \ref{fig:operator-plan} shows the AgentSpeak plan that the operator executes upon every new simulation step. This plan introduces the operator's desire to process the agent identification notifications (!processFriendlies), monitor the current task assignments (!updateTaskAssignments), and assign new tasks to free agents (!assignTasks).

\subsubsection{Task Assignment}
The operator maintains a set of task assignment beliefs to keep track of each builder's current task assignment. Each belief takes on the form $taskAssignment(Agent, Task, Req)$ where Agent is the builder agent name, Task is the assigned task name, and Req is a structure containing information about the assigned task requirement. On top of keeping track of the current task assignments, the operator also uses these beliefs to determine the free agents --- the builder agents that are not in any current task assignment beliefs. As demonstrated by the structure of the belief, and as we will discuss, a builder will only be responsible for one requirement block for a given task. 

The process of task assignment involves selecting a task and assigning it to a sub-team of free agents. The operator calls an internal action to help with the process of task selection and assignment. Since the internal action has access to the information in the agent containers, it can utilize the container information to determine the free agents that can communicate and collaborate on a task. The internal action iterates through the list of free agents provided by the operator, and checks if any of them have mutually identified one another. Sub-teams are then created based off of the mutually-identified agents, this guarantees that the agents are able to communicate and utilize the connect action with any other member of the sub-team. The internal action will attempt to maximize the number of agents in a sub-team, allowing it to take on larger tasks.

Once the sub-teams are determined, the internal action decides on a task assignment for the sub-team and assigns one requirement of that task to each sub-team member. Since each sub-team agent is responsible for only one requirement block, the sub-team can only be assigned a task that has a number of requirements less than or equal to the number of sub-team members. If a task is assigned to a sub-team, but has less requirements than the sub-team has members, any builders that are not assigned a requirement will remain free agents and will be removed from the sub-team. 

If there are no free agents that have identified one another, no sub-teams or tasks will be assigned to the agents. It is also possible for multiple sub-teams to be formed, each will have their own mutually-identified members and unique task assignment. Once the internal action processes all of the free agents, it will return a list of all of the task and requirement assignments back to the operator. The operator will process each of the task and requirement assignments, and will add new task assignment beliefs to its belief base while also notifying any builders of their new task (and requirement) assignments.

\subsubsection{Task Monitoring}
Before performing the task assignment process, the operator will first update its list of current task assignments. The operator monitors each task assignment by iterating through each assignment belief and ensuring the assigned task is still valid. The task is considered valid if: 1. the task has not expired, and 2. the task has not yet been submitted by any team. If these conditions are not met, the operator will remove any task assignments associated with the invalid task and will notify any of the affected builders.

\subsection{Shared Strategy Components}
Despite the differences in strategy and behaviour, the attackers and builders share common elements and lower-level strategies in order to achieve their respective goals. Both roles, for example, will utilize some form of movement at some point in their lifetime, and it would be favourable to share some of the contingency plans associated with movement failure. It therefore makes sense to design and organize the AgentSpeak code in a way that maximizes the amount of code that can be utilized by both roles. Both roles share the same code base, the difference is that they have different goals that they aim to achieve.

\subsubsection{Agent Identification}
Both roles perform agent identification in order to share map knowledge, however, agent identification is crucial for builders as they require it to collaborate on tasks. The identification process is the same for all agents. Builders can identify and share information with the attackers, and vice-versa. Since we run the reasoning cycle asynchronously from the simulation cycle, this allows the agents to reason about multiple intentions, making it possible to handle and process multiple possible events such as agent identification and agent navigation, within one simulation step.

\subsubsection{Navigation and Exploration}
The navigation and exploration components provide the agent with plans for handling map navigation to a destination, searching for map things (dispensers, goal spaces, etc.), and intelligent map exploration. These components require access to the map model stored in the agent container object, therefore they all utilize internal actions to obtain their required information. 

When searching for map things, the agent utilizes an internal action to query the map model for a specified thing. The internal action will return the virtual absolute location of the closest map cell that matches the request of the agent (for example, a dispenser), otherwise the internal action fails to unify if nothing in the map model matches the request. If nothing is found, the agent will explore the map until the internal action can resolve the request. If the map model can resolve the request and a virtual location is unified, the agent will utilize \textit{destination navigation} to get to the unified destination.

The agents have the ability to explore the map in an intelligent way. Rather than aimlessly exploring in any given direction, the agents can utilize the map model to determine the closest missing chunk of the map. The internal action for exploration unifies the best direction to explore in, in order to lead the agent to unexplored territory. This means that when an agent is exploring, they will almost always find new areas of the map with undiscovered things and terrain. Exploration is the default behaviour for all agents, they will explore the map until they have received enough information to achieve their respective goals.

Destination navigation utilizes path-finding on the map model to determine the shortest path from the agent's current location to a specified destination. The internal action will unify the path as a list of cardinal directions (N, S, E, W) that the agent must navigate in order to get to the destination. If the destination does not exist on the map model, or if there is no path to get to the destination, the destination navigation plan will fail. Jason will propagate this failure up the intention stack until a contingency is found to handle the failure. In cases where this happens, the agent typically will end up exploring until a path is available, but it is ultimately up to the implemented contingency plans to decide how to handle this failure.

If a path to the destination exists, the agent will follow the navigation path provided. If the next direction in the provided path is blocked (either by an obstacle or map things), or if the destination is not what the agent expected (the map got regenerated and caused the map model to be out-of-date), then the agent will call the path-finding internal action again. The internal action will determine the next best path to the destination, or will fail if none exists.

\subsubsection{Action Plans and Contingency}
Each scenario action (move, connect, attach, etc.) is managed by a plan in Jason. Each of these action plans have contingency plans in the case of any failures. They also provide some level of failure avoidance, implemented with respect to the action being performed. 

To demonstrate this with the move action: if an agent were to attempt to move west, but one of its attachments is blocking movement, the move action plan will detect this and attempt to find a rotation where moving west is possible before it attempts the move action. 

\begin{figure}[h]
\centering
\includegraphics[width=\textwidth]{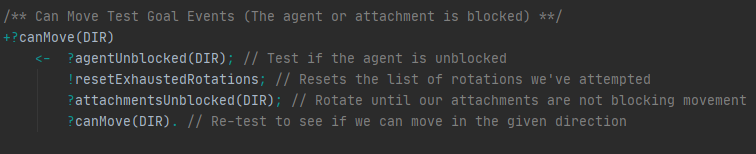}
\caption{The canMove test goal allows the agent to exhaust the various rotations until we are no longer blocked by our attachments.}
\label{fig:canMove-test}
\end{figure}

The move plan utilizes the canMove test goal shown in Figure \ref{fig:canMove-test} to rotate if our attachments are blocking movement. This test goal plan will rotate until the movement direction is no longer blocked. Once unblocked, the agent then performs the original move request. 

Test goals are used in this way to bring the agent to a desirable and expected state in order to avoid failures. If a rotation can not be found and the movement can not be unblocked, the plan will fail and will propagate up to any parent intentions or contingency plans to handle. 

By avoiding and handling failures in this manner, we try to minimize the number of steps that the agents waste by sending actions that are guaranteed to fail or by being in a bad state. Jason's ability to manage failures through contingency plans definitely proved itself useful in this area, allowing us to capture and recover from various edge cases and bad states.

\subsubsection{Getting Unstuck}
While developing the agents, it was a common occurrence for the agents to be spawned in a location that was surrounded by obstacles. Without any mechanism for breaking out of this sort of situation, the agents essentially become useless for the rest of the simulation (unless they get lucky and an opposing agent or clear event clears up the obstacles for them). It is also possible for an agent to be blocked by things, such as another entity or a block. Both the attacker and builder agents utilize the same strategy in order to free themselves.

\begin{figure}[h]
\centering
\includegraphics[width=0.75\textwidth]{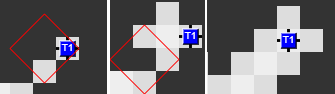}
\caption{The agent utilizes clear actions to break out of a contained space. The agent clears the obstacles that are most likely to open it up to free space.}
\label{fig:getting-unstuck}
\end{figure}

The agents will utilize an internal action in order to determine whether or not they are blocked in all directions (or ``caged in"). The internal action will then determine the best obstacle to clear, which is determined as the obstacle with the highest number of surrounding free-spaces. By following this criterion, the agent can attempt to clear out a path that requires a minimal number of clear actions. Figure \ref{fig:getting-unstuck} demonstrates the strategy that the agents use to break out of their contained area.

\subsection{Attacker Strategy}
The attackers explore the map and utilize the clear action in order to interfere with the other team. When starting off, the attackers will explore the map until they discover any goal terrain. The attackers will then monitor the goal clusters by navigating to a known goal terrain location (including goal locations shared by the map knowledge of other agents). The attackers will look for opposing agents residing on a goal terrain that have at least one block attached to them, and will then use the clear action targeted at one of the attached blocks in an attempt to reject the opponent's potential task submission.

\begin{figure}[h]
\centering
\includegraphics[width=\textwidth]{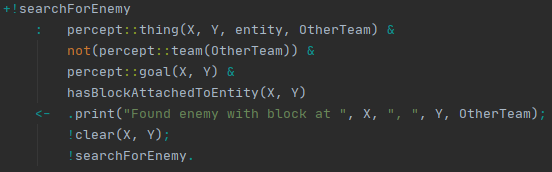}
\caption{The plan responsible for attacking enemies. This plan is considered once the attacker detects an enemy with a block attachment that resides on a goal cell.}
\label{fig:attacker-plan}
\end{figure}

\begin{figure}[h]
\centering
\includegraphics[width=0.5\textwidth]{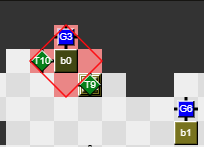}
\caption{Attacker agents T9 and T10 attacking opponent G3's block during simulation 1 of the match against GOAL-DTU.}
\label{fig:attacker-example}
\end{figure}

Figure \ref{fig:attacker-plan} shows the clear action being executed by the attacker agent when an enemy is detected on a goal cell with an attached block. Figure \ref{fig:attacker-example} shows attacker agents T9 and T10 attacking opponent G3 in the first simulation against GOAL-DTU.

\subsection{Builder Strategy}
The builders are responsible for performing the main scenario objective of requirement gathering, task building, and task submission. The builder agents separate this into multiple sub-goals. First, the builders explore the map while waiting for a task requirement assignment from the operator. After receiving an assignment, the builders will obtain their respective requirement blocks. Following this, the task's master builder determines a meeting point and communicates with each individual slave builder in order to sequentially deliver and connect each task requirement at the meeting point. Once all of the task requirements are connected, the master builder must submit the task. 

Each of the sub-goals associated with fulfilling the task requirements have their own AgentSpeak plans. Each sub-goal plan must make sure that the builder successfully achieves the objective of the sub-goal. These plans also have their own respective contingency plans for managing and correcting any sub-goal-specific failures. If any of these sub-goal failures can not be managed or corrected by any of the contingency plans, the agent will detach any blocks and will re-attempt its current task requirement assignment from the start.

\subsubsection{Free Agents: Waiting For a Task Assignment}
When a builder is a free agent, it will wait for a task requirement from the operator. The builder has no other option but to explore the map in order to be somewhat productive, since it can't start working on a task yet. Exploring the map allows the builders to build on their current knowledge of the map while also identifying other agents on the team. As the number of mutually-identified free agents increases, the operator will be able to assign larger tasks to the sub-teams. By having a larger sub-team, the chances of being assigned a task will increase since you will be considered for a larger set of tasks.

\subsubsection{Obtaining Block Requirements}
When a new task assignment has been received, the builder will detach any existing attachments that are not relevant to the current requirement block type. When performing this, the builder will make sure it is outside of a goal cluster before detaching any blocks in order to prevent the detached block from cluttering up the goal space as a courtesy to both teams. 

After detaching all irrelevant blocks, the builder will first check to see if it has the required block type already attached. If it does, the agent moves on to the next sub-goal (delivering the block). Otherwise, the agent must find a dispenser to obtain the block. 

\begin{figure}[h!]
\centering
\includegraphics[width=\textwidth]{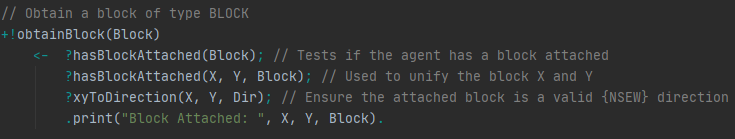}
\caption{The obtainBlock plan used for obtaining a specific block type. The plan utilizes test goals to ensure a predictable state before continuing onto the next step of the strategy.}
\label{fig:obtain-block}
\end{figure}

\begin{figure}[h!]
\centering
\includegraphics[width=\textwidth]{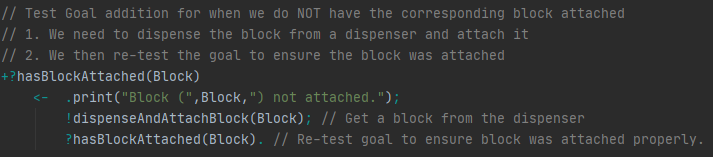}
\caption{The test goal plan for when a block is not currently attached to the agent. The agent will search for the appropriate block dispenser, request a new block, and then attach it to itself.}
\label{fig:notAttached}
\end{figure}

Figure \ref{fig:obtain-block} shows the obtainBlock plan for obtaining a given block type. Figure \ref{fig:notAttached} shows the test goal plan that gets executed when the agent does not have the given block type attached to itself. This test goal is used by obtainBlock in order to ensure a block is attached before continuing.

When searching for the dispenser, the builder calls an internal action that provides the closest dispenser that dispenses the necessary block type. If the appropriate dispenser can not be found, the agent will explore until map knowledge of the required dispenser is introduced. When the appropriate dispenser is found, the agent will navigate to the dispenser, request the block from the dispenser, and attach the dispensed block to itself.

\subsubsection{Delivering Blocks and Builder Coordination}
Once the builders have the required block attached to themselves, they deliver the block to a common meeting point and connect their blocks to the other builders working on the same task. The sub-team responsible for each task will have one master builder, and multiple slave builders. 

\begin{figure}[h!]
\centering
\includegraphics[width=0.8\textwidth]{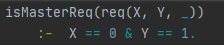}
\caption{The AgentSpeak rule responsible for determining if the current agent's task assignment permits them to be the master builder.}
\label{fig:isMaster-rule}
\end{figure}

The master builder is the builder who was assigned the first ordered requirement --- the requirement with a location of (0, 1). The AgentSpeak rule responsible for determining this is provided in Figure \ref{fig:isMaster-rule}.

The master builder determines the meeting point for block delivery and connection for the sub-team, and also synchronizes the task requirement sequence with each slave builder. The slave builders --- the builders responsible for all requirements that do not have a location of (0, 1) --- will take turns delivering their attached blocks to the master builder. The master builder first navigates to the determined \textit{meeting point}. The meeting point is where the slave agents will deliver and connect their block requirements with the master builder. The meeting point will always be on a goal terrain, and will be selected such that there is enough space below the master builder to connect the blocks from each slave. 

\begin{figure}[h!]
\centering
\includegraphics[width=\textwidth]{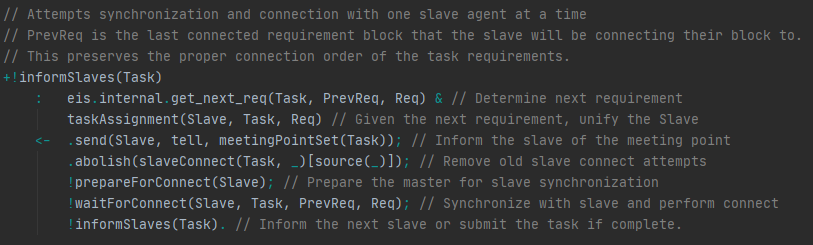}
\caption{The plan responsible for helping the master builder coordinate with each of the slave builders.}
\label{fig:informSlaves-Master}
\end{figure}

Figure \ref{fig:informSlaves-Master} demonstrates the plan that allows the master builder to coordinate with each slave. The master builder will coordinate the connection of the requirement blocks with the slaves. It utilizes the task requirement planning algorithm to determine the next slave that needs to deliver and connect their assigned requirement. The master builder will then notify the next slave builder that it must deliver and connect the requirement to the master, providing the slave with the destination location for the block. Slave builders that have attached their requirement block, but have not yet been notified by the master builder to deliver their requirement, will explore the map until they are notified.

\subsubsection{Block Connection}
\begin{figure}[h!]
\centering
\includegraphics[width=\textwidth]{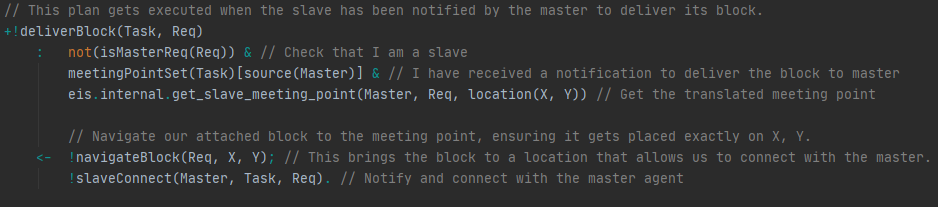}
\caption{The plan responsible for coordinating the slave builders with the master builder.}
\label{fig:deliverBlock-Slave}
\end{figure}

As each slave gets notified to deliver their respective requirement block, they will navigate and rotate themselves in a way that ensures that the block is on the exact destination cell that the master builder requested. Once the block is on the specific destination, the slave builder will notify the master builder of the successful delivery. The master will then notify the slave that it is ready to connect, and both agents will attempt to connect their blocks. This behaviour is demonstrated by the AgentSpeak plan shown in Figure \ref{fig:deliverBlock-Slave}.

If the connect action fails unexpectedly, both agents will try the connection again on the next step. Upon successful block connection, the slave agent will detach itself from the connected block. The block delivery and connection process is then repeated with the slave builder responsible for the next requirement in the task. If all requirements have been connected, the master builder will proceed to task submission.

\subsubsection{Task Submission}
\begin{figure}[h!]
\centering
\includegraphics[width=\textwidth]{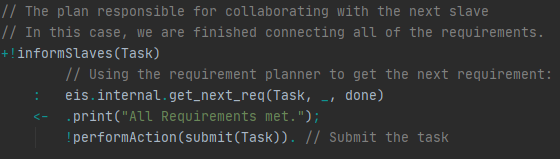}
\caption{When the master goes to inform the next slave, the context of this plan allows us to handle the case where we are done building the task. Since we are already on a goal space, we attempt task submission.}
\label{fig:taskSubmit}
\end{figure}

Since the master has already predetermined the meeting point of block delivery to be on a goal cell, the master can submit the task immediately after connecting all of the requirements with the slave builders. Once the master submits the task successfully, it will notify the operator. The operator will remove the relevant task and requirement assignment beliefs for the sub-team. Figure \ref{fig:taskSubmit} shows the plan responsible for task submission.

In the next section, we discuss the performance of the TRG agents in the competition. We will use various metrics to help measure the effectiveness of each of the strategies discussed in this section, providing insight on what aspects of our team strategy went well and what could have been improved.

% Agent Performance
\section{Agent Performance}
The TRG team won 1 simulation, tied 2, and lost 6. In this section we analyze some of the match statistics, providing some speculation about the issues encountered by the agents during the competition and where things could have been improved. The data used for the metrics in this section were obtained from the competition match replays.

\subsection{Match Performance Metrics}
The performance of the team throughout the various matches can be examined through a total of 10 different metrics. Each of the metrics can be used to measure the performance of at least one aspect the agent strategy.

\subsubsection{Metric 1: Attachment Utilization}
The attachment utilization metric is used to show how many blocks were obtained, and how many of those obtained blocks were actually used in tasks. The only reason that the builders attach blocks is so that they can complete tasks. When this metric is low, this signifies that the builders are wasting time obtaining and attaching blocks. This metric is provided as the number of used attachments / the number of obtained attachments.

\subsubsection{Metric 2: Number of Connections Made}
The number of connections made is provided as a way to measure the number of successful block connections between builders. Successful connections demonstrate that the agents are able to collaborate correctly, this further measures the ability to perform block delivery and connection synchronization between the master and slave builders.

\subsubsection{Metric 3: Submitted Tasks}
This metric is the number of submitted tasks in the simulation. This measures the overall ability of the builders.

\subsubsection{Metric 4: Failed Submissions}
This is the number of failed submissions. A failed submission occurs when the agents gather and connect all of the block requirements, but fail to submit a task. This metric can signify issues with the builder task submission process, or an issue with the builder's ability to complete tasks before the deadline.

\subsubsection{Metric 5: First Task Start Time}
The first task start time is defined as the simulation step of the first block attachment used in the first task submission. This metric measures how quickly the agents can explore the map to identify one another, get assigned a task from the operator, and then attach the first task requirement block. This shows how long the agents take before getting starting on task building.

\subsubsection{Metric 6: Average Task Requirement Size}
This is the average number of requirements for all of the tasks submitted in a simulation. This measures the team's willingness and ability to complete larger tasks.

\subsubsection{Metric 7: Average Task Completion Time (Per Requirement)}
The average task completion time measures the team's ability to complete tasks in a timely manner. This is calculated for each task by measuring the amount of time between the first block attachment, and the task submission time. This result is then divided by the number of task requirements so that the completion time is normalized to account for the task size.

\subsubsection{Metric 8: Average Attach to Connect Time}
After attaching a block to itself, the slave builders must then connect the block to the master builder. This metric, which measures the average time between block attachment and connection, shows how long the slave builders take to deliver and connect their blocks with the master builder. This metric is directly influenced by the distance between the location of the block dispenser --- where the slave agent attaches the block requirement --- and the meeting location set by the master agent --- where blocks are delivered and connected. This metric is improved by choosing a meeting point close to the necessary block dispensers. 

\subsubsection{Metric 9: Average Last Connect to Submit Time}
The average time between the last successful block connection and the task submission time is used to measure the amount of time the master builder spends getting to a goal space and submitting the task. As we will see in the match performance results, the master builders maintained a consistently low value for this metric. This is attributed to the meeting point selection algorithm, which always brings the master builder to a goal cell before the requirements are delivered and connected by the slave agents. Upon successfully building the task requirements, the master builder can immediately submit the task.

\subsubsection{Metric 10: Opponent Rejected Submissions}
This single metric is used to measure the performance of the attacker agents. This metric defines a ``rejected submission" as a successful clear operation on an opposing agent who is located on a goal space with an attached block. The clear action will successfully disable the agent and will remove any of its attached blocks affected by the clear action. 

For the clear action to be performed, it must be executed 3 steps in a row on the same space. Even if an opposing agent did not have the intention of task submission, we assume that since the agent was waiting with a block on a goal cell, it was somehow involved in the task submission process. By disabling the opposing agent and removing at least one of its attached blocks, we count this as a rejected submission.

\subsection{General Remarks and Improvements}
Before diving into the details about the performance of the agents in each match, we would like to first discuss some of the general short-comings that were noticed during review of all simulation replays. 

\subsubsection{Task Failures}
There were a few cases where builders gathered and constructed the requirements for a given task, but missed the deadline within a few steps. This occurred in simulation 3 against LFC (missed 40 points for task14), and simulation 2 against FIT BUT (missed 90 points for task1). In simulation 3 of the LFC match, one of the builders attempted to submit a task outside of a goal cluster. As expected, the submission failed and the master builder ended up dropping the task. This could have been easily prevented if a contingency plan existed for the failure of task submission resulting from being outside of a goal space. In this case, the contingency plan could have been implemented such that it brings the master builder to a goal space and re-attempts submission, rather than just dropping the task.

\subsubsection{Task Collaboration Bug}
The task collaboration bug occurred as a result of an issue with the planning algorithm. This resulted in the master builder not communicating properly with any of the slave builders after the second requirement; this caused both the master and slave builders assigned to the given task to wait indefinitely, or at least until the task became invalid. This would only occur with tasks that had more than 2 requirements. This issue was eventually fixed mid-way through the first simulation of the last team match, and unfortunately cost the agents a lot of points (and matches).

The team struggled to complete tasks in all simulations of the matches against GOAL-DTU and LFC, and the first simulation against FIT BUT. Fortunately, the bug was fixed, and the agents were able to improve their match performance for the last two simulations. This helped the agents score one win, and although they lost their last simulation, they were able to score a large amount of points.

\subsubsection{Task Follow-through}
The builders seemed to attach a large amount of blocks from the dispensers, but did not connect or submit these attachments as part of a task. This implies that the builders wasted a lot of time dispensing and attaching blocks just to detach them later on. This premature detachment of blocks resulted from various things, including the inability of slave builders to line up with the task master due to blocked paths, and exceeding task deadlines and being assigned a new task, among other things. 

The underlying reason for this is that the agents typically detach a block if they are stuck or in a bad state. Anything that results in this state, including getting stuck in a crowded area or failing an action too many times can trigger the agent to reset itself back to a known state, which includes detaching any blocks. A better way to approach this would be to have a higher tolerance for failure before resetting the state, or to improve detection and understanding of the current surroundings. For example, we can account for agent density in an area and attempt to force movement through the usage of the clear action before requiring a state reset.

\subsubsection{Late Task Start Time}
For the majority of the matches, the first requirement of the first task was not attached to the agent until about half-way through the match. This metric shows how long it takes for the builders to explore the map and identify each other before they are assigned a task by the operator. Although this is heavily influenced by map size and initial placement of the agents, the agents did not get started on any tasks until around step 217 on average. 

\subsubsection{The Attacker Agents}
The attacker agents will be evaluated based off of the number of opposing task submissions they reject or interrupt. The attacker agents did not have any coordination strategy, and as a result some of the attackers --- in some cases, all of the attackers --- would monitor the same empty goal cluster. In these cases, we have agents that are essentially sitting at a goal cluster and doing nothing at all.

In the cases where the attacker agents do interrupt any opposing agents, the number of submission rejections is very low. In the best case scenario (simulation 3 against FIT BUT), the attackers were able to reject 5 task submissions. In most simulations, the combined rejection rate of all of the attackers is at most 1 task rejection. The performance of the attackers was extremely poor. 

In retrospect, not much time was put into the attacker strategy. It would have been smarter to either assign a smaller portion of the team to attackers, to just get rid of them completely, or to have all agents be builders and allow them to dynamically switch their role to an attacker when they are not busy with any tasks.

\subsection{Match Performance Results}
The match performance of the TRG team is analyzed in detail utilizing the previously defined match performance metrics. Each match will be discussed briefly, and will be accompanied with a table demonstrating the results of each simulation with respect to the metrics. 

Each of the simulation column headings will be colour-coded to dictate whether or not the TRG team won (green), tied (yellow), or lost (red), the corresponding simulation. Task metrics will be greyed out if they are not applicable to the simulation --- meaning the builders did not submit any tasks.

\subsubsection{Match 1 Performance (vs. GOAL-DTU)}
TRG struggled to submit tasks in all three simulations against GOAL-DTU. After reviewing the replays, the builders seemed to struggle with block delivery and navigating to their meeting points after gathering their respective requirements. This left the builder agents wandering around the map with attached blocks until their tasks expired. Although the attacker agents were able to reject a few submissions from the other team, it was not enough to compensate for the poor performance of the builders.

The builders were able to submit a total of 1 task with 2 requirements across all 3 simulations. This scored us a grand total of 40 points against GOAL-DTU, who was able to score a total of 120 points. The metrics for this match can be seen in Table \ref{tab:dtu-results}.

\begin{table}[h!]
\centering
\resizebox{\textwidth}{!}{%
\begin{tabular}{|l|c|c|c|}
\hline
\multicolumn{1}{|c|}{\textbf{Metrics (vs. GOAL-DTU)}} &
  \multicolumn{1}{l|}{\cellcolor[HTML]{FFFFC7}\textbf{Simulation 1}} &
  \multicolumn{1}{l|}{\cellcolor[HTML]{FFCCC9}\textbf{Simulation 2}} &
  \multicolumn{1}{l|}{\cellcolor[HTML]{FFCCC9}\textbf{Simulation 3}} \\ \hhline{====}
\textbf{Score (GOAL-DTU - TRG)} & 40 - 40 & 40 - 0                   & 40 - 0                  \\ \hline
\textbf{Attachment Utilization (Used/Obtained)} & 2 / 27 & 0 / 29                   & 0 / 33                   \\ \hline
\textbf{Number of Connections Made}          & 1      & 0                        & 0                        \\ \hline
\textbf{Submitted Tasks}                     & 1      & 0                        & 0                        \\ \hline
\textbf{Failed Submissions}                  & 0      & 0                        & 0                        \\ \hhline{====}

\textbf{First Task Start Time}               & 253    & \cellcolor[HTML]{C0C0C0} & \cellcolor[HTML]{C0C0C0} \\ \hline
\textbf{Avg. Task Requirement Size}          & 2      & \cellcolor[HTML]{C0C0C0} & \cellcolor[HTML]{C0C0C0} \\ \hline
\textbf{Avg. Task Completion Time (Per Req.)}           & 40     & \cellcolor[HTML]{C0C0C0} & \cellcolor[HTML]{C0C0C0} \\ \hline
\textbf{Avg. Attach to Connect Time}    & 72     & \cellcolor[HTML]{C0C0C0} & \cellcolor[HTML]{C0C0C0} \\ \hline
\textbf{Avg. Last Connect to Submit Time}         & 2      & \cellcolor[HTML]{C0C0C0} & \cellcolor[HTML]{C0C0C0} \\ \hhline{====}

\textbf{Opponent Rejected Submissions}             & 2      & 3                        & 1                        \\ \hline
\end{tabular}%
}
% \vspace{1mm}
\caption{The agent metrics for simulations 1-3 against GOAL-DTU. All time metric values are provided in steps.}
\label{tab:dtu-results}
\vspace{-8mm}
\end{table}

\subsubsection{Match 2 Performance (vs. LFC)}
Our agents were able to perform marginally better in the second match against LFC. Although the builder agents finally got their act together and starting building tasks correctly, they unfortunately were not able to meet a few of the task deadlines, resulting in failed submissions. On top of this, the builders also struggled to build any tasks that had more than 2 requirements. This was a direct result of the task collaboration bug.

It is safe to say that the attacker agents were completely useless in all three simulations, rejecting a total of 1 submission from the other team. This match could have benefited from less attackers and more builders. The final score across all three simulations was 390 (LFC) - 160 (TRG). The metrics for the match against LFC can be seen in Table \ref{tab:lfc-results}.

\begin{table}[h!]
\centering
\resizebox{\textwidth}{!}{%
\begin{tabular}{|l|c|c|c|}
\hline
\multicolumn{1}{|c|}{\textbf{Metrics (vs. LFC)}} &
  \multicolumn{1}{l|}{\cellcolor[HTML]{FFCCC9}\textbf{Simulation 1}} &
  \multicolumn{1}{l|}{\cellcolor[HTML]{FFFFC7}\textbf{Simulation 2}} &
  \multicolumn{1}{l|}{\cellcolor[HTML]{FFCCC9}\textbf{Simulation 3}} \\ \hline
\textbf{Score (LFC - TRG)} & 180 - 120 & 0 - 0                   & 210 - 40                 \\ \hline
\textbf{Attachment Utilization (Used/Obtained)} & 6 / 26 & 0 / 20                   & 2 / 23 \\ \hline
\textbf{Number of Connections Made}          & 3      & 1                        & 4      \\ \hline
\textbf{Submitted Tasks}                     & 3      & 0                        & 1      \\ \hline
\textbf{Failed Submissions}                  & 0      & 1                        & 2      \\ \hhline{====}

\textbf{First Task Start Time}               & 225    & \cellcolor[HTML]{C0C0C0} & 326    \\ \hline
\textbf{Avg. Task Requirement Size}          & 2      & \cellcolor[HTML]{C0C0C0} & 2      \\ \hline
\textbf{Avg. Task Completion Time (Per Req.)}           & 19.2   & \cellcolor[HTML]{C0C0C0} & 16     \\ \hline
\textbf{Avg. Attach to Connect Time}    & 25.7   & \cellcolor[HTML]{C0C0C0} & 21     \\ \hline
\textbf{Avg. Last Connect to Submit Time}         & 2      & \cellcolor[HTML]{C0C0C0} & 2      \\ \hhline{====}

\textbf{Opponent Rejected Submissions}             & 0      & 0                        & 1      \\ \hline
\end{tabular}%
}
% \vspace{1mm}
\caption{The agent metrics for simulations 1-3 against LFC. All time metric values are provided in steps.}
\label{tab:lfc-results}
\vspace{-5mm}
\end{table}
\subsubsection{Match 3 Performance (vs. FIT BUT)}
After finally tracking down the task collaboration bug mid-way through simulation 1, the builders were able to significantly improve their task performance in the two remaining simulations. TRG secured their first (and only) win in the second simulation against FIT BUT. In the simulation, the builders were able secure the win by submitting 4 tasks, including a task with 3 requirements. 

Unfortunately, the same can not be said about the attackers. Their performance was abysmal in the first two simulations, and although they made a small impact in simulation 3, it was not enough to prevent FIT BUT from scoring 500 points in the simulation. The total score across all 3 simulations was 660 (FIT BUT) - 390 (TRG). The metrics for the match against FIT BUT can be seen in Table \ref{tab:fit-results}.

\begin{table}[h!]
\centering
\resizebox{\textwidth}{!}{%
\begin{tabular}{|l|c|c|c|}
\hline
\multicolumn{1}{|c|}{\textbf{Metrics (vs. FIT BUT)}} &
  \multicolumn{1}{l|}{\cellcolor[HTML]{FFCCC9}\textbf{Simulation 1}} &
  \multicolumn{1}{l|}{\cellcolor[HTML]{9AFF99}\textbf{Simulation 2}} &
  \multicolumn{1}{l|}{\cellcolor[HTML]{FFCCC9}\textbf{Simulation 3}} \\ \hline
\textbf{Score (FIT BUT - TRG)} & 80 - 0 & 80 - 210                   & 500 - 180                  \\ \hline
\textbf{Attachment Utilization (Used/Obtained)} & 0 / 26                   & 9 / 29 & 6 / 43 \\ \hline
\textbf{Number of Connections Made}          & 1                        & 6      & 4      \\ \hline
\textbf{Submitted Tasks}                     & 0                        & 4      & 2      \\ \hline
\textbf{Failed Submissions}                  & 1                        & 1      & 0      \\ \hhline{====}

\textbf{First Task Start Time}               & \cellcolor[HTML]{C0C0C0} & 112    & 167    \\ \hline
\textbf{Avg. Task Requirement Size}          & \cellcolor[HTML]{C0C0C0} & 2.25   & 3      \\ \hline
\textbf{Avg. Task Completion Time (Per Req.)}           & \cellcolor[HTML]{C0C0C0} & 27.2   & 23.7   \\ \hline
\textbf{Avg. Attach to Connect Time}         & \cellcolor[HTML]{C0C0C0} & 49.5  & 37.3  \\ \hline
\textbf{Avg. Last Connect to Submit Time}         & \cellcolor[HTML]{C0C0C0} & 2      & 2      \\ \hhline{====}

\textbf{Opponent Rejected Submissions}             & 0                        & 0      & 5      \\ \hline
\end{tabular}%
}
% \vspace{1mm}
\caption{The agent metrics for simulations 1-3 against FIT BUT. All time metric values are provided in steps.}
\label{tab:fit-results}
\vspace{-12mm}
\end{table}

\section{Conclusion}
This paper presented the approach taken by the TRG agents, detailing the design and implementation of the agents, and the system architecture that provides the backbone for reliable reasoning and behaviour within the competition. The high-level system components that are implemented in Java provide information processing abilities that abstract away from the agents and the AgentSpeak code. The agents access these components through the usage of internal actions and rely on them to provide a useful representation so that the agents may fulfill their intentions as reliably and as accurately as possible.

The high-level system components, which includes the navigation system, the agent identification system, and the requirement planner, among others, all work together to help each agent achieve their respective goals (such as attacking the opposing team, task requirement gathering, collaboration with other team agents, and task submission). The agents are assigned one of two roles and communicate with a centralized agent --- the operator --- who coordinates the identification of agents, and assigns tasks to builder agents so that they may work together and complete tasks in the simulation. 

This paper also presented the various debugging challenges that were faced during the course of agent development. Various approaches were taken in order to improve the overall agent debugging experience. The best approach was to use a combination of the various tools available for debugging, which includes the custom-developed visualization tool, IntelliJ IDE support, and the Jason-provided agent mind inspector. These tools all provide differing levels of information about the agents, and combining their usage with one another provided the most informative and efficient debugging approach.

Every aspect of agent development brought on new and interesting challenges. As we attacked each challenge, and watched the system and agent behaviour evolve, we also re-evaluated the purpose of each system component, as well as the system as a whole. We went through various iterations of strategy and system design which eventually led up to the multi-agent system detailed in this paper.

Considering the challenges we faced during design and development of the agents, and during the competition, we were very happy with the performance of the agents. Even though the team performed poorly overall in the contest, it was very fulfilling to witness the system's components and the agent behaviour work together as they competed against other agents. Although the attacker agents were basically useless, and we could have performed better with a team full of builders, it was still interesting and fun to develop a strategy for the attackers just to see how they would affect the other teams.

We will briefly discuss the limitations of our multi-agent system, and any future work that needs to be done. The agent visualization tool that was created for the purpose of the TRG agents is tightly-coupled to the structure of the agent containers. Improvements to the visualization tool would be to provide a generic interface available to any agent. This would allow the visualization tool to be used by any multi-agent system, and could ideally improve the process of debugging agents for other developers. Additional work could be done to integrate this tool with the Jason mind inspector.

Our multi-agent system also has the drawback of having design components that are tightly-coupled to the MAPC scenario. For example, the agent container and the parsed percept objects are all derived from scenario-specific concepts and challenges. Further work needs to be done on this aspect of the system to generalize these components. These components are meant to handle challenges specific to the MAPC scenario, but if generalized, could be applied to address similar design concerns and challenges faced by other multi-agent systems.

In terms of team strategy, further work could be done to mitigate some of the shortcomings faced during the competition. The attacker role would be reconsidered to minimize the loss of potential task submissions due to an ineffective attacking strategy. The builder role could also be further improved by having multiple master builders within a task sub-team. This allows the builders to work on subsections of the assigned task. This results in a decentralized approach to task building, and would significantly reduce the amount of time that the builders spend waiting for each slave to connect with a single master. 

We hope our design, debugging, and implementation endeavours can help provide some insight into some of the problems that other multi-agent systems face. As a final note, we would like to thank the MAPC organizers for hosting the competition, and for providing such a challenging, yet fulfilling, scenario for the agents.

% Competition Questionnaire
\newcommand{\bolditem}[1]{\item \vskip0.5em\textbf{#1}}

\section{Team overview: short answers}
\subsection{Participants and their background}

\begin{description}
\bolditem{What was your motivation to participate in the contest?}
The main motivation was to use the simulation as a concrete example for a thesis topic; while also allowing us to explore the benefits and limitations of Jason.

\bolditem{What is the history of your group? (course project, thesis, $\ldots$)}
Babak is Michael's thesis supervisor. He has helped tremendously with regards to the development of the agents, and the paper.

\bolditem{What is your field of research? Which work therein is related?}
We aim to explore the intersection of multi-agent systems and epistemic logic. This contest provides many challenges that could be approached with epistemic logic. Although this was not attempted during the contest, we plan on using the simulation as a means for researching the formulation of (and potential solution to) these challenges using epistemic logic.
\end{description}

\subsection{Statistics}
\begin{description}
\bolditem{How much time did you invest in the contest (for programming, organizing your group, other)?}
An average of 40-50 hours a week was spent programming from mid-July to October. No time was spent organizing the group.

\bolditem{How many lines of code did you produce for your final agent team?}
Total AgentSpeak LOC: 2309
Total Java LOC: 7385

\bolditem{How many people were involved?}
Only one of the authors (Michael) was directly involved in the agent programming; although weekly meetings were made with Babak as he was able to provide some higher-level direction regarding some of the challenges that were faced in this contest. Babak also helped immensely with regards to the structuring and revision of the content in this paper.

\bolditem{When did you start working on your agents?}
Agent development started in May 2019. Initially, it involved just playing around with the simulator and attempting to understand the ``Agents Assemble" scenario. Serious agent development did not start until mid-July; continuing until the contest in October.
\end{description}

\subsection{Agent system details}
\label{sec:strategies}

\begin{description}
\bolditem{How does the team work together? (i.e. coordination, information sharing, ...) How decentralized is your approach?}
The team works together through both information sharing and coordination. The agents are able to share perceived map information with each other. Agents who are assigned to the same task will coordinate and synchronize with one another, when it is necessary to connect the requirements. There is an external agent, referred to as the operator. The operator does not participate directly in the simulation, but rather exists as a central location for processing team information and notifications. In general, the agents operate independently of each other until it is necessary to coordinate. 

\bolditem{Do your agents make use of the following features: Planning, Learning, Organizations, Norms? If so, please elaborate briefly.}
The agents plan the task requirements using an in-house algorithm that follows the task generation algorithm of the simulation server. In terms of organization, there are two main roles that an agent may take on, which directly influences the behaviour of the agent.

\bolditem{Can your agents change their behaviour during runtime? If so, what triggers the changes?}
The behaviour of an agent depends on the designated role, and the current state of the agent. The role of an agent is completely static and does not change during runtime. However, an agent that has the responsibility of completing tasks may change its behaviour depending on what the current goal is (obtaining blocks, navigating, exploring, etc.).

\bolditem{Did you have to make changes to the team (e.g. fix critical bugs) during the contest?}
Yes, I had discovered an issue with the planning algorithm; this resulted in the agents not connecting the blocks properly. Unfortunately, this issue was fixed too late into the simulation.

\bolditem{How did you go about debugging your system?}
Debugging the behaviour of the agents was the most difficult and time-consuming aspect of the competition. A lot of effort was made to ease the process of debugging, and I found myself relying on Java (especially the internal actions) to compensate for my lack of Jason knowledge and understanding. Jason does come equipped with the agent mind inspector, although I found that I commonly encountered the observer effect while attempting to debug the behaviour of the agents. In order to make the debugging process as painless as possible, we combined the usage of the mind inspector with IDE breakpoints, and a custom-designed visualization tool.

\bolditem{During the contest you were not allowed to watch the matches. How did you understand what your team of agents was doing? Did this understanding help you to improve your team's performance?}
Through the usage of the custom-designed agent visualization tool, I was able to partially observe the match. The tool provides a visual representation of the map model shared by the agents, and also provides some debugging information. This tool allowed me to visualize what my own team was doing, but didn't provide much insight about the opposing team.

\bolditem{Did you invest time in making your agents more robust? How?}

After the initial qualification round, I had spent the next two full weeks rebuilding my agents. Some components were completely rebuilt from the ground-up. One of the main issues I was facing before the rebuild, was the synchronization of the map between the identified agents. Perceptions were being written to the map model incorrectly, and providing incorrect data to the path finding algorithm. Upon careful reconstruction of the code base, the agents were able to correctly read and update the shared map model. The rebuilding of the system led to the system design detailed in this paper.

\end{description}

\subsection{Scenario and Strategy}
\begin{description}

\bolditem{What is the main strategy of your agent team?}
The main strategy of the team is split between two roles. The attackers, which is composed of half of the team, are responsible for chasing down any opponents and using the clear action to destroy any of their blocks.

The other half of the team is composed of the builder agents who aim to complete tasks. The builders will explore the map, until they are assigned a task and requirement by the operator agent. Once they are assigned a task and a requirement, they obtain the required block from a dispenser, and then meet at a designated meeting location. They utilize the connect action to connect their block requirements, and then submit the task. The coordination and building of task requirements is done by a master builder.

There is also an operator agent who is responsible for centralized processing of team information.

\bolditem{Your agents only got local perceptions of the whole scenario. Did your agents try to build a global view of the scenario for a specific purpose? If so, describe it briefly.}

Yes, the agents maintained and shared information contained within their respective map models. Each agent's map model is built using the map perceptions the agent receives at the start of every step. The agents can request information from this map model.

\bolditem{How do your agents decide which tasks to complete?}

In order for the agents to communicate and collaborate, they must first identify each other. Once a set of agents are identified by one another, they are then able to collaborate on a task. The agents are then assigned a task based on the number of agents in the identified set (the sub-team), the number of requirements needed, and the deadline. The agents prioritize tasks with a higher number of requirements and a further deadline. 

\bolditem{Do your agents form ad-hoc teams to complete a task?}
Yes, tasks are assigned to sub-teams. The sub-teams consist of the agents who have mutually identified one another.

\bolditem{Which aspect(s) of the scenario did you find particularly challenging?}
The most difficult aspect was trying to get the agents to correctly time the connection of blocks, without causing them to block if the connect action didn't go according to plan. The most time-consuming aspect (although, not the most difficult) was working with the limited perception range.

\bolditem{If another developer needs to integrate your techniques into their code (i.e., same programming language tools), how easy is it to make that integration work?}
\end{description}
I attempted to design the code base with modifiability and readability in mind, but as the contest got closer these qualities started to become less of a priority. It is quite a large code base, and I attempted to document it as best as I could, so it shouldn't be too difficult to understand what the agents are doing.

\subsection{And the moral of it is \ldots}
\begin{description}
\bolditem{What did you learn from participating in the contest?}
The contest helped me with learning and understanding the Jason agent environment (including understanding AgentSpeak, using the provided agent debugger, etc.).

\bolditem{What are the strong and weak points of your team?}
The agents were designed with a lot of contingency plans in mind. In the case of a failed action, or if the current perceptions are different than what the agent would have expected, the agents are able to adjust their current goals so that they can be in a state that allows them to continue on with what they need to be doing. 

Unfortunately, having a lot of contingency plans drastically increased the size of the code base, and also made the behaviour of the agents very difficult to debug. Another weak point of the team is that they rely heavily on the map model to perform path finding. Although the agents are resilient to unexpected changes in the map, the limited perception range makes it extremely difficult for agents to reliably navigate to certain destinations.

\bolditem{Where did you benefit from your chosen programming language, methodology, tools, and algorithms?}
I'd say the main benefit that I gained from my choice of language, was the ability to call Java code from Jason (through internal actions) as it helped compensate for my lack of Jason knowledge, also helping immensely with any debugging issues I had.

\bolditem{Which problems did you encounter because of your chosen technologies?}
The biggest problem I encountered was losing a lot of time due to not being able to efficiently debug any agent behaviour issues.

\bolditem{Did you encounter new problems during the contest?}
Yes, I discovered a few issues with the task requirement planning algorithm. Although this technically was not a new problem, it was an existing problem that decided to express itself and hinder the performance of the agents during the competition.

\bolditem{Did playing against other agent teams bring about new insights on your own agents?}
Other than the fact that our team's attacker agents were useless, there wasn't much new insight brought on from the other teams.

\bolditem{What would you improve (wrt. your agents) if you wanted to participate in the same contest a week from now (or next year)?}

The one thing I would do to improve my agents is to develop debugging tools from the very beginning. It was difficult to properly trace a bug using the Jason debugging tools, and so I had to develop my own tool to help understand what any of my agents were doing. If I had developed this tool from the start, it would have greatly improved development and debugging efficiency.

\bolditem{Which aspect of your team cost you the most time?}

I'd say that debugging unwanted behaviour was the most time-consuming task. Typically, any issues that occurred in my AgentSpeak code took at least 2-3 days to debug and fix. As time went on, the Jason code became more complex, and because of that it was difficult to pin-point the source of a bug.

\bolditem{What can be improved regarding the contest/scenario for next year?}

The contest was really well organized and the documentation was done well. It would be nice to have a skip action from the start, as it saves a lot of time, especially while debugging. It would also be convenient if the simulation provided a way to parse the raw percepts from the EnvironmentInterface into percept-specific Java objects.

\bolditem{Why did your team perform as it did? Why did the other teams perform better/worse than you did?}
I think the main reason for performing poorly in this contest was that I split the team in half. Rather than having half of the team attack the other team, I think I should have dedicated all agents to completing tasks. On top of this, my agents had an issue with planning their requirements during the contest. This caused the agents to drop their blocks prematurely and go back to re-obtain the block through the dispenser. This drastically reduced the number of tasks that were submitted.

\end{description}

\bibliography{./references}
\bibliographystyle{ieeetr}

\end{document}